\documentclass[conference]{IEEEtran}
\IEEEoverridecommandlockouts
\usepackage{cite}
\usepackage{amsmath,amssymb,amsfonts}
\usepackage{algorithmic}
\usepackage{graphicx}
\usepackage{textcomp}
\usepackage{xcolor}
\usepackage{censor}
\usepackage{subcaption}
\usepackage{booktabs}
\usepackage{tabularx}
\usepackage[a4paper, total={184mm,239mm}]{geometry}
\def\BibTeX{{\rm B\kern-.05em{\sc i\kern-.025em b}\kern-.08em
    T\kern-.1667em\lower.7ex\hbox{E}\kern-.125emX}}

\begin{document}

\title{Active Inference on the Edge: A Design Study*
\thanks{* Funded by the European Union (TEADAL, 101070186). Views and opinions expressed are however those of the author(s) only and do not necessarily reflect those of the European Union. Neither the European Union nor the granting authority can be held responsible for them.}
}



\author{
Boris Sedlak, Victor Casamayor Pujol, Praveen~Kumar~Donta, and Schahram Dustdar\\[0.2cm]

\IEEEauthorblockA{\textit{Distributed Systems Group}, 
    Vienna University of Technology (\textit{TU Wien}), Vienna 1040, Austria.\\   Email: \{b.sedlak, v.casamayor, pdonta, dustdar\}@dsg.tuwien.ac.at\\
    }}

\maketitle

\begin{abstract}
Machine Learning (ML) is a common tool to interpret and predict the behavior of distributed computing systems, e.g., to optimize the task distribution between devices. As more and more data is created by Internet of Things (IoT) devices, data processing and ML training are carried out by edge devices in close proximity. To ensure Quality of Service (QoS) throughout these operations, systems are supervised and dynamically adapted with the help of ML. However, as long as ML models are not retrained, they fail to capture gradual shifts in the variable distribution, leading to an inaccurate view of the system state. Moreover, as the prediction accuracy decreases, the reporting device should actively resolve uncertainties to improve the model's precision. Such a level of self-determination could be provided by Active Inference (ACI) -- a concept from neuroscience that describes how the brain constantly predicts and evaluates sensory information to decrease long-term surprise. We encompassed these concepts in a single action-perception cycle, which we implemented for distributed agents in a smart manufacturing use case. As a result, we showed how our ACI agent was able to quickly and traceably solve an optimization problem while fulfilling QoS requirements.
\end{abstract}


\section{Introduction}


Recent years have reported a constant transition of logic from the central cloud towards the edge of the network \cite{deng_edge_2020}, thus, closer to the Internet of Things (IoT) devices that actually generate data. This transition includes the training of Machine Learning (ML) models (i.e., to save bandwidth and improve privacy), as well as data processing (i.e., to decrease latency) \cite{pujol2023edge}. As soon as training has finished, ML models are a common measure to interpret and predict the behavior of distributed systems, e.g., to estimate the impact of redeployment \cite{chen_causeinfer_2019} or forecast potential system failures \cite{morichetta_demystifying_2023}, which must be circumvented to ensure the Quality of Service (QoS).

ML models are applied throughout the Computing Continuum (CC), i.e., from the cloud, over the fog, to the network edge -- close to where models were trained. However, in many cases, ML models are not retrained, although new observations would be available \cite{morichetta_demystifying_2023,chen_causeinfer_2019}; this inevitably leads to an inaccurate view of the system state, which, in turn, decreases the quality of any inference mechanism that uses the ML model. Imagine an elastic computing system, like envisioned in \cite{sedlak_controlling_2023,nastic_sloc_2020}, which observes the system through a set of metrics, evaluates whether QoS requirements -- also called Service Level Objectives (SLOs) -- were fulfilled, and dynamically reconfigures the system to ensure SLOs are met. If the variable distribution changes and the ML model is not adjusted, this makes it impossible to interpret the metrics correctly, and any consequential reconfiguration will fail to fulfill its purpose.

Ensuring the precision of an ML model requires a continuous feedback mechanism; such behavior could, for example, be achieved by optimizing a value function, as in reinforcement learning \cite{martinez_probabilistic_2021,friston_reinforcement_2009}. However, we believe that this requires a more holistic approach, which starts with making the SLOs first-class citizens during the ML training process. Further, we want to highlight the responsibility of any service that uses the ML model to actively resolve or report ambiguities. Such a level of self-determination could be provided by Active Inference (ACI), a concept from neuroscience that describes how the brain constantly predicts and evaluates sensory information to decrease long-term surprise. ACI combines various concepts that have already been rudimentarily implemented in distributed systems, e.g., causal inference to identify dependencies between system parts \cite{chen_causeinfer_2019}, or dynamic adaptations of the system to ensure QoS -- called homeostasis. This shows the potential of ACI.


In this paper, we advance one step further by combining the ACI concepts in a comprehensive design study of an ACI agent that optimizes the throughput of a smart factory. Internally, the agent follows an action-perception cycle: First, it estimates which parameter assignments would violate given SLOs, then it compares this expectation with new observations, and finally, it adjusts its beliefs (i.e., the ML model) accordingly. The agent focuses on exploring values that promise a high throughput while avoiding such that are likely to violate the SLOs. Furthermore, it favors solutions that are likely to improve the model precision, which, in turn, provides the agent with a clear understanding of the causal relations between model variables. Hence, the contributions of this article are the following:

\begin{itemize}
    \item A novel ML paradigm based on ACI that continuously evaluates the quality of predictions. Thus, agents improve the model precision to ensure QoS for ongoing operations.
    \item The composite representation of an agent's behavior throughout the action-perception cycles. The distinct factors can be fine-tuned to determine the agent's preferences.
    \item A complete design study for a smart manufacturing agent that paves the way for other researchers to implement ACI in related automative use cases.
\end{itemize}

The remainder of the paper is structured as follows: Section~\ref{sec:background} provides background information on ACI principles in distributed systems; in Section~\ref{sec:related-work} we present existing work that included ACI; within Section~\ref{sec:design-process} we outline the design process of an ACI agent, which we implemented and evaluated in Section~\ref{sec:analysis}. Finally, Section~\ref{sec:conclusion} concludes the paper.


\section{Background}
\label{sec:background}

We consider ACI an unknown concept for most readers outside of neuroscience; therefore, we use this section to summarize core concepts of ACI according to Friston et al.~\cite{friston_life_2013,kirchhoff_markov_2018,friston_reinforcement_2009,smith_step-by-step_2022,sajid_active_2021}. This includes but is not limited to (1) free energy minimization, (2) hierarchical organization of beliefs, (3) action-perception cycles, and (4) Bayesian inference and belief updating. Following that, we delineate our view of the intersection between ACI and distributed systems, in particular edge computing. 

\subsection{Active Inference Principles}

To interpret observable processes, agents construct generative models, e.g., a person would reason that it rains due to water drops falling from the sky. 
Based on these observations, the agent can learn to understand real-world processes. However, if the generative model and the process diverge, the agent will eventually be ``surprised", e.g., because water drops were caused by a neighbor watering her plants.
The discrepancy (or uncertainty) between the agent's understanding of the process and reality is called Free Energy (FE); a more accurate understanding decreases FE at the same time.

More formally, the surprise $\Im (o|m)$ of observation $o$ given model $m$ is the negative log-likelihood of the observation. The surprise itself is capped by the FE of the model -- expressed as the Kullback-Leibler divergence ($D_{K\!L}$) between approximate posterior probability ($Q$) of the hidden states ($x$) and their exact posterior probability ($P$). While mathematical approaches, such as exemplified in Eq. \eqref{eq:surprise} \& \eqref{eq:free-energy}, provide a much-needed notation for working with the FE principle, in practice, many variables are computationally intractable, e.g., the true probability $P$.


\begin{equation}
    \Im (o|m)= -\ln\!\!{\overbrace{P(o|m)}^\text{Model Evidence}}
    \label{eq:surprise}
\end{equation}
\vspace{-12pt}
\begin{equation}
    F[Q,o] = \underbrace{D_{K\!L}[Q(x)||P(x|o,m)] + \Im (o|m)}_\textrm{(Variational) Free Energy} \geq \Im (o|m) 
    \label{eq:free-energy}
\end{equation}

\vspace{5pt}

Internally, agents organize their generative models in hierarchical structures; each level interprets lower-level causes and, based on that, provides predictions to higher levels. For example, suppose it rains with a certain probability, I bring an umbrella. This is commonly known as Bayesian inference and allows agents to use existing beliefs (widely known as priors) to calculate the probability of related events. Such decision processes can be segregated into self-contained causal structures (i.e., Markov blankets), e.g., one to interpret the weather and another to dress. As the agent infers that it is raining, he decides to pick the umbrella. However, when dressing, the agent only considers the weather (\textit{rainy} or \textit{sunny}) and disregards lower-level observations that led him to conclude that it's raining (e.g., humidity or obscurity).

ACI agents constantly engage in action-perception cycles, where they (1) predict sensory inputs, actively seek the information, and update their beliefs depending on the outcome -- widely known as predictive coding. 
Afterward, they (2) can adjust the world to their existing beliefs through their own actions. 
While pragmatic actions (e.g., picking an umbrella) fulfill agents' preferences (e.g., staying dry), agents improve their decision-making by exploring the environment through epistemic actions. For example, a mere look at the sky reveals that the neighbor has watered her plants, avoiding surprise when wrongfully leaving with an umbrella. The agent thus updates its prior beliefs (i.e., rain $\rightarrow$ water) according to new data (i.e., rain $\rightarrow$ water $\leftarrow$ flowers) to form posterior beliefs.

\subsection{ACI Principles in Distributed Systems}
\label{subsec:aci-principles}


ACI encompasses multiple concepts; although there exist few implementations that combine them in one framework, most of them can be encountered in distributed systems. In the following, we review the principles described above and map them to existing concepts as far as possible:

\subsubsection{Causal Inference}

Causal structures (e.g., Bayesian networks~\cite{pearl_probabilistic_1988}) can be trained to identify dependencies between parts of distributed systems. As pointed out by \cite{ganguly_review_2023_short}, causal structures have the fundamental advantage (over deep learning) of justifying their actions or recommendations, thus improving trustworthiness. Distributed systems can explain how metrics (e.g., latency or CPU load) are related to the system state~\cite{sedlak_controlling_2023},
backtrack which service or device caused a system failure~\cite{chen_causeinfer_2019}, or predict the impact of redeployment~\cite{tariq_answering_2008}.

\subsubsection{Free Energy Minimization}

AI models are trained to improve their prediction accuracy, which, in turn, reduces their FE. Energy-based models \cite{teh_energy-based_2003}, in particular, rate uncertainty as (free) energy. To ensure model accuracy over time, one option is to continuously report prediction errors (e.g., \cite{veeramachaneni_ai2_2016}). However, in many cases, systems lack adequate feedback loops and thus fail to capture gradual shifts in the variable distributions (e.g., \cite{tariq_answering_2008,chen_causeinfer_2019,simsek_detecting_2020})
While ML training is essential to decrease FE, epistemic actions often suffice to reduce uncertainties about expected outcomes: Distributed systems resolve contextual information before executing pragmatic actions, e.g., identify a low-utilized agent for load balancing or task offloading \cite{huang_vehicle_2020,guo_toward_2020}, or evaluate resource availability before scaling a system \cite{sedlak_controlling_2023,furst_elastic_2018}. It is the general tradeoff between seeking either pragmatic value (exploitation) or information (exploration); multi-agent systems (e.g., \cite{levchuk_active_2019}) control this through a hyperparameter called ``exploration rate", which fosters early exploration of a global value space but decays over time as agents report little improvement.
To improve generative models whenever feasible, this is also implemented for edge-based systems \cite{huang_vehicle_2020}.


\subsubsection{Homeostasis}

The ultimate goal for an ACI agent is to persist over time; this requires maintaining certain internal variables under control. This concept is called homeostasis and can be found in various systems: The human body, for example, requires a core temperature of 37° for chemical processes; distributed systems, on the other hand, specify QoS requirements as SLOs \cite{furst_elastic_2018,nastic_sloc_2020,pusztai_novel_2021,sedlak_controlling_2023}. While the human body has its own temperature-controlling mechanisms, distributed systems rely on elasticity strategies to ensure QoS, e.g., by scaling computational resources to cap response time. Although surprise plays a significant role here, e.g., when reporting SLO violations, the preferred strategy is to engage with the environment to correct this instead of changing the perception.

\section{Related Work}
\label{sec:related-work}

While, to the best of our knowledge, there exists no complete implementation of ACI in distributed systems; a handful of research works have combined ACI with computer science: 

The authors in \cite{vilas_active_2022} discuss ACI as a general computational framework, highlighting how existing research used ACI for (simulating) sensory processing. Touching on the design of ACI agents, Heins et al.~\cite{heins_pymdp_2022} provide a Python simulation that exemplifies how to structure action-perception cycles. Heins et al. further remark that existing ACI research largely focuses on formally constructing models in isolated environments such as Matlab SPM (e.g., \cite{smith_step-by-step_2022}) rather than putting them into action, e.g., to improve the precision of ML models. A more hands-on application of ACI is thus to extend reinforcement learning with ACI principles
\cite{martinez_probabilistic_2021,friston_reinforcement_2009}. However, most research to date either uses only a few ACI principles or is not applied enough to easily transfer presented concepts to distributed systems.


The work in \cite{levchuk_active_2019} is, therefore, an exception because it embeds ACI into the IoT and describes how ACI can improve the behavior of adaptive agents. Thus, individual agents may dynamically regroup into hierarchical teams, federate knowledge, and collectively strive after a common goal (i.e., a search task). By emphasizing the exchange of experiences between agents, they were able to speed up the convergence of the distributed task. However, while they focused on FE minimization, they did not treat the other two principles we identified for ACI in distributed systems: causal inference and homeostasis. In this paper, we will present an agent that uses all three ACI principles to infer actions, maintain agents' internal equilibrium, and persist over time. Nevertheless, we will use the representation from \cite{levchuk_active_2019} for FE minimization.
\\





\section{Active Inference Design Process}
\label{sec:design-process}

In the following, we will walk through the design of an ACI agent by (1) building upon ACI background information to draw an action-perception loop, (2) describing a use case where the agent trains a model from scratch to optimize performance, (3) marking the boundaries of the generative model trained, and (4) defining the agent's behavior throughout the cycles.

\subsection{Action-Perception Loop}

To continuously ensure the precision of ML models and any consequential action, we will employ self-evidenced agents, i.e., they reason about their environment and train models autonomously. To that extent, ACI agents operate in action-perception cycles; each iteration aims to improve the accuracy of the model, infer optimal actions, and thus persist over time. 
As such, agents can be embedded into distributed systems, e.g., to maintain the QoS for a distributed task.

\begin{figure}[t]
\centerline{\includegraphics[width=0.50\textwidth]{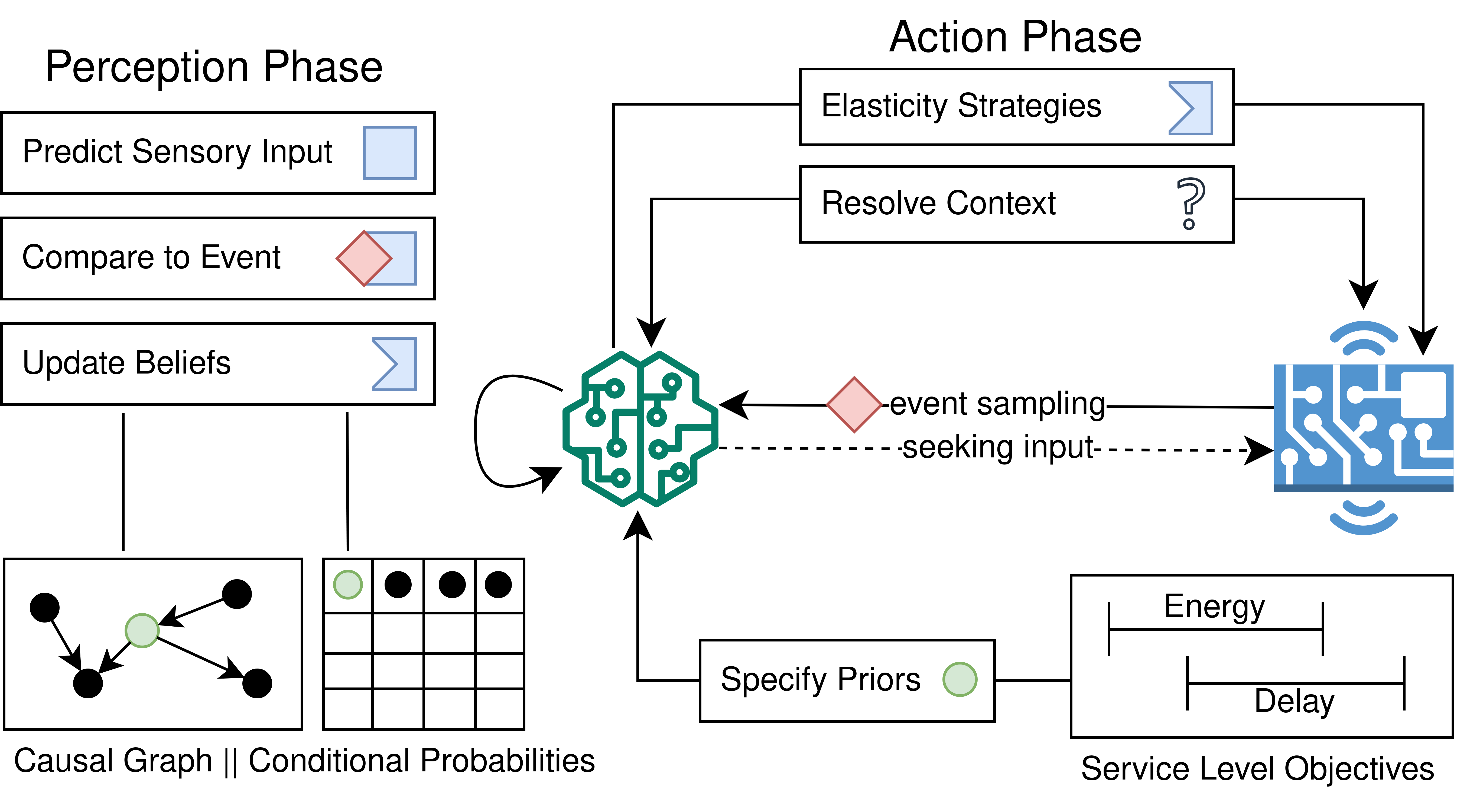}}
\caption{Overview of the action-perception cycle in ACI}
\label{fig:action-perception}
\end{figure}

Fig.~\ref{fig:action-perception} provides a high-level overview of the steps that are repeated by the agent:
Initially, a set of SLOs define the agent's preferences (e.g., $delay \leq \delta$) and establishes its expectations prior to evaluating any sensory data. The agent then assembles a causal graph to determine which factors influence these parameters; the conditional probability table contains the degree to which they are affected. Afterward, the agent starts to continuously predict the probability of observations, might actively seek a corresponding input, and then compares the event against the expectation. To decrease FE, the agent now has three options: (1) adjust its beliefs accordingly, i.e., update the causal graph and conditional probabilities; (2) change the environment toward its preferences, e.g., executing elasticity strategies; or (3) resolve contextual information to improve decision-making.

\subsection{Use Case Description}

\begin{figure}[t]
\centerline{\includegraphics[width=0.48\textwidth]{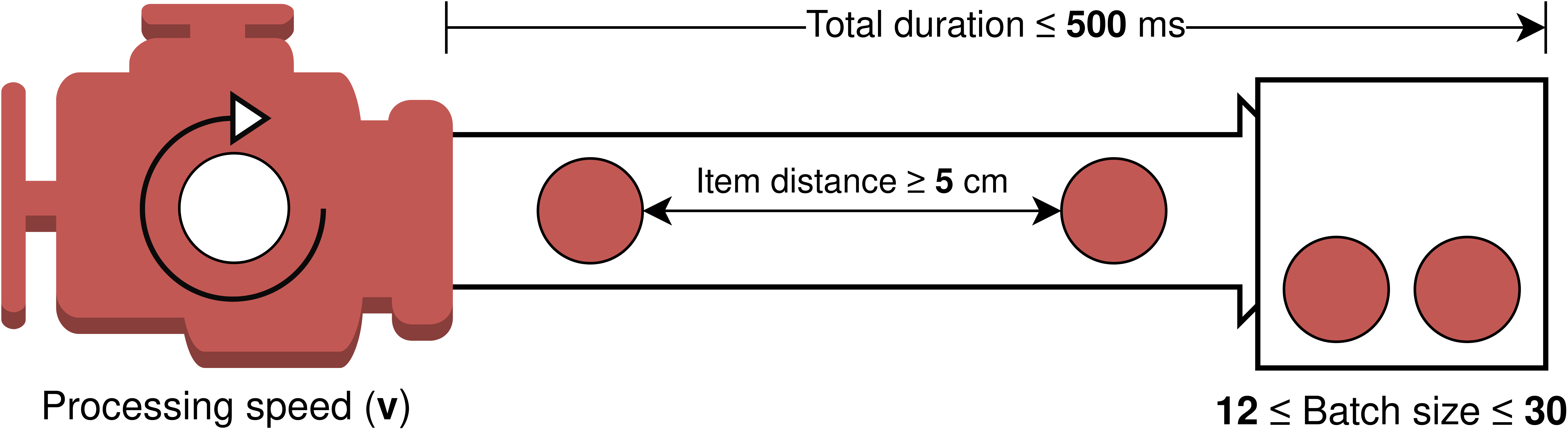}}
\caption{A factory producing machine parts in batches}
\label{fig:use-case}
\end{figure}

The following use case is embedded in the smart manufacturing environment, which provides numerous opportunities for sensor-oriented analysis and dynamic adaptation of production. Fig.~\ref{fig:use-case} provides a high-level overview of the use case:

Within a factory, machine parts are fabricated in batches of 12 to 30 pieces; a larger batch size increases the throughput and utilization of the factory engine. Each batch must be completed within 500 ms; thus, an increasing batch size decreases the timeframe for processing each part.
The engine's utilization is supposed to impact the processing duration, though the magnitude is unknown. Also, due to a consecutive assembly step, the distance between parts should be above 5 cm.
Given this setup, the factory manager would like to know the largest batch size that fulfills all constraints.
However, because they lack historical data, it is difficult to answer this by training an ML model; this issue must be actively explored. Ideally, the learning process would also be autonomous and allow the factory to simultaneously produce their goods.

To provide the factory manager with the optimal batch size, we supervise the factory engine through an ACI agent. The resulting smart engine now enters what can be understood as a continuous calibration mechanism: throughout its action-perception cycle, it (1) estimates if an increase or decrease in batch size would violate the given constraints (i.e., its SLOs), (2) compares the expectation with the result, and (3) continuously explores the value space by slightly varying the batch size.
The agent thus gradually approaches solutions that promise high throughput while satisfying all constraints.


\subsection{Generative Model Setup}

While the use case showed how ACI can help solve optimization problems, we will now dive deeper into the generative model created by the agent. The design process is loosely oriented towards the guidance provided by Parr et al.~\cite{parr_active_2022}, which depicts an abstract sequence of steps to design ACI systems. The main questions we aim to answer are:
\vspace{5pt}

\begin{enumerate}
    \item What is part of the generative model, and what are the interfaces to the exterior?
    \item What is the hierarchical and temporal depth of the model, and how do they affect causal inference?
    \item What are the model variables and prior beliefs -- what can be modified (i.e., learned), and what is immutable?
\end{enumerate}
\vspace{8pt}

\begin{table}[t!]
  \centering
  \caption{Model variables and their boundaries}
  \label{tab:model-variables}
  \begin{tabular}{lclc}
    \toprule
    Name   & Unit & Description &  Range  \\
    \midrule
    \textit{batch size}           & num        & number of machine parts per batch  & $[12, 30]$\\
    \textit{utilization}          & \%        & utilization of the factory engine  & $[1, 100]$ \\
    \textit{distance}             & cm        & space between two machine parts  & $[1, \infty[$\\
    \textit{part delay}          & ms        & processing time per machine part  & $[1, \infty[$ \\
    \textit{batch delay}          & ms        & total time for batch processing  & $[1, \infty[$ \\
    \bottomrule
  \end{tabular}
\end{table}

\begin{figure}[t]
\centerline{\includegraphics[width=0.46\textwidth]{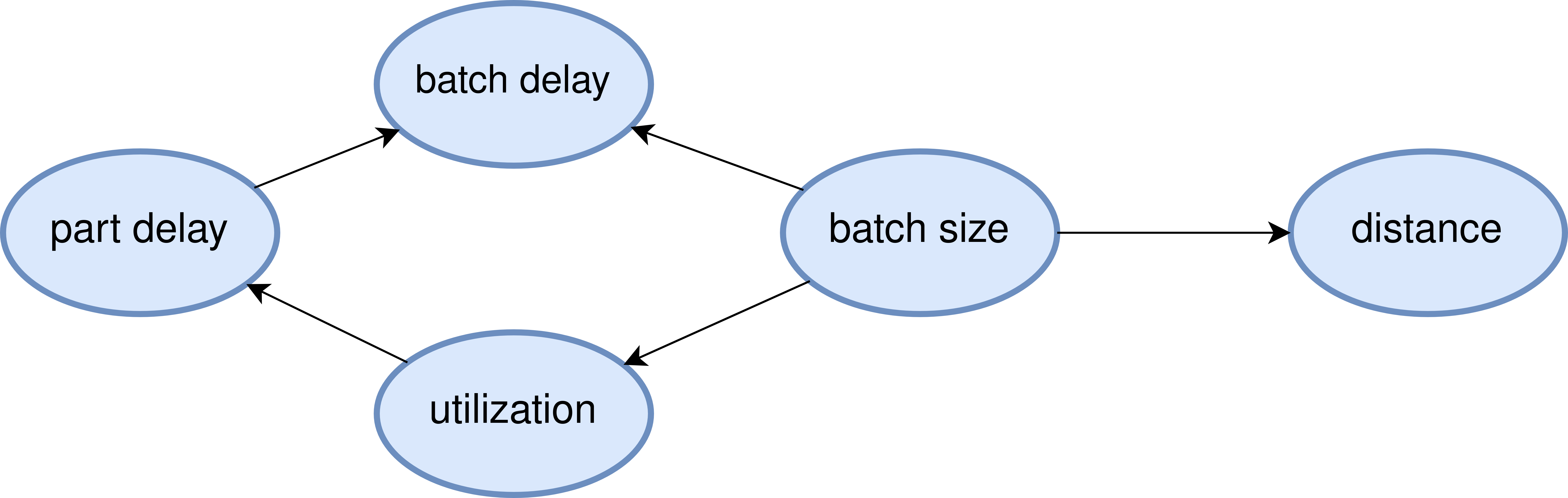}}
\caption{Initial beliefs of relations between model variables}
\label{fig:dag}
\end{figure}

To predict whether a \textit{batch size} would fulfill the SLOs, agents must identify the variables that have an impact on them.
These could be extracted through a causal structure (e.g., \cite{tariq_answering_2008,chen_causeinfer_2019})
or, in the absence of training data, come from expert knowledge, which can be updated over time. The manager initially believes that variables are related as depicted by the Directed Acyclic Graph (DAG) in Fig.~\ref{fig:dag}; the respective variables are described in Table~\ref{tab:model-variables}. Variables in ACI represent an interface between the generative model and the exterior world, i.e., if the utilization of the physical engine changes, this is reflected through the respective variable (i.e., \textit{utilization}), which in turn determines the internal view of the system state.
%
Information provided through interface variables is used to construct the generative model, but also to evaluate SLO fulfillment (e.g., \textit{batch delay} $\leq$ 500 ms). To that extent, it needs to analyze the respective variable (from the DAG), as well as its parent, child, and spouse nodes. This subset provides a causal filter to the variable state, called Markov blanket. A central premise is that all these sensory variables accurately reflect the exterior; otherwise, subsequent decisions (e.g., decreasing \textit{batch size} to decrease \textit{delay}) would perpetuate any measurement error.

For the given use case, we use an SLO-induced boundary as our natural limit on temporal depth: Equal to the maximum \textit{batch delay (bd)}, each action-perception cycle lasts 500 ms. Within each cycle, the agent predicts the engine's behavior (i.e., reflected through the metrics) over the next 500 ms; afterward, the prediction is compared against the events observed during that time.
While the cycle's length can be chosen freely, longer periods decrease the prediction accuracy or increase the computational complexity (i.e., to evaluate the SLO once, it must consider multiple cycles or fractions of them). The hierarchical depth, on the other hand, is determined by the number of variables and edges in the model. A deeper hierarchy would increase the complexity of model training and inference; however, the use case does not provide variables other than the ones already contained in the DAG.

So far, it only remains to explain what priors are in the given example: Priors are our assumptions about the system before verifying them, e.g., which \textit{batch size} should provide the highest throughput without violating the SLOs. Priors are subject to the learning process, while SLOs are fixed; each action-perception cycle aims to improve the generative model's accuracy, thus decreasing FE.
As we will see in Section~\ref{sec:analysis}, the initial beliefs (i.e., before evaluating any cycle) speed up the convergence of ACI to the optimal solution.

\subsection{Active Inference Agent}

To find the optimal \textit{batch size}, the central mechanism of the agent is the action-perception cycle shown in Fig.~\ref{fig:action-perception}. Initially, the agent has little information available to form priors or infer a fitting \textit{batch size}; however, as the agent samples the environment through its interface variables, each cycle adds new observations ($s_n$) to the total amount of known samples ($s_k$). The agent's behavior throughout each cycle (i.e., how it interprets sensory information and which action it takes) is determined by three main factors: (1) pragmatic value ($pv$) of actions; (2) ambiguity or risk assigned ($ra$) to actions; and (3) epistemic value or information gained ($ig$) by actions. The following notation of these factors is related to \cite{levchuk_active_2019,parr_active_2022}, though the composition is different.
The only parameter that the agent can actively set is \textit{batch size}; the remaining variables are causally influenced by this factor. Thus, if the agent changes the \textit{batch size}, this is reflected through the related variables.

The pragmatic value that emerges from higher \textit{batch size} is simple: more throughput. Therefore, we define $pv (bs) = bs \times \frac{100}{30}$, which encourages the agent to increase \textit{batch size}. The multiplier $\frac{100}{30}$ scales the factor to the range $[1,100]$, which is equal for all three factors.
Contrarily, high \textit{batch size} might exceed $bd \leq 500 ms$ or $d \geq 5 cm$, i.e., the SLOs associated with \textit{batch delay} and \textit{distance} ($d$). To evaluate the risk of violating the SLOs we consider how often past observations for a \textit{batch size} ($s_{kb}$) have violated the SLOs. The $ra$, e.g., for $batchsize=20$, would thus be determined by the rate between samples that fulfilled the SLOs and the total number of samples ($|s_{kb}|$); this is formalized in Eq. \eqref{eq:ra} \& \eqref{eq:valid}. As long as the list of samples for a $batchsize$ (or short $bs$) is empty (i.e., $|s_{kb}| = 0$), the agent interpolates the value with the prior and latter $ra$ as reference points, e.g., if the agent knows $ra(30)=90$ and $ra(20)=20$, in the absence of samples for $batchsize=25$, it infers that $ra(25)=55$. This interpolation is contained in Eq.~\eqref{eq:inter}.
\vspace{4pt}
\begin{equation}
        ra (bs) = 100 -
\begin{cases}
inter(bs), & \text{if }|s_{kb}|=0
\vspace{5pt}\\
\frac{valid (bs)}{|s_{kb}|} \times 100, & \text{otherwise}
\end{cases}
\label{eq:ra}
\end{equation}

\vspace{-4pt}

\begin{equation}
inter(bs) = ra_{i-1} + (bs - bs_{i-1}) \times \frac{(ra_{i+1} - ra_{i-1})}{(bs_{i+1} - bs_{i-1})}
\label{eq:inter}
\end{equation}

\vspace{-4pt}

\begin{equation}
    valid (bs) = \sum_{i=1}^{|s_{kb}|} [(bd_i \leq 500) \land (d_i \geq 5)]
    \label{eq:valid}
\end{equation}

The $ig$ of an action is determined by the ambiguity that it resolves; in other words, we aim to make future predictions less surprising. Reviving the idea of surprise from Eq. \eqref{eq:surprise}, we now require the surprise for $s_n$ given $s_k$: Eq. \eqref{eq:surprise-batch} shows how the total surprise is the sum of surprises of new samples; $f(x)$ describes the probability density function\footnote{A function that described the likelihood of an observation $o$ in a continuous range given that the probabilities are distributed with $O \sim \mathcal{N}(\mu, \sigma)$.} with $\mu = \bar{s_k}$ and $\sigma_{s_k}$. For each $s_n$, the surprise is appended to a list of past surprises $S = S \cup surprise (s_n,s_k)$; $S_x \in S$ contains all values with $x = batchsize$. If a $batchsize$ has reported repeatedly surprising values, it supposedly provides more information gain: This is reflected through Eq. \eqref{eq:ig} because the median surprise ($\tilde{S}_x$) will rise above the global average ($\bar{S}$). To foster exploration of prior unknown $batchsize$, in the absence of surprise values, e.g., $|S_{25}| = 0$, it assumes $ra(25) = \max(S)$.
\begin{equation}
    ig (bs) = \left(\frac{\tilde{S}_{bs}}{\bar{S}}\right) \times 100
    \label{eq:ig}
\end{equation}

\vspace{-10pt}

\begin{equation}
    surprise (s_n,s_k) = \sum_{i=1}^{|s_{k}|} - \log f(d_i)
    \label{eq:surprise-batch}
\end{equation}

Ultimately, to evaluate the potentials but also risks that emerge from each $batchsize$, the three factors are merged into a common one -- ($cf$). Since all factors are scaled to the range $[1,100]$, they can be combined as $cf(bs) = pv(bs) - ra (bs) + ig (bs)$. At the end of each cycle, the agent resolves $cf(x)$ for $x=[12,30]$ and chooses the highest scoring as new $batchsize$.

\section{Evaluation}
\label{sec:analysis}

To evaluate the ideas presented in the last Section, we provide a Python implementation of the ACI agent that comprises the action-perception loop to create a generative model. Although we did not embed the agent in a physical engine to measure sensory information, we used a compatible data set generated with \cite{sedlak_privacy_2023} to simulate an equal behavior. The prototype of the agent, the data, as well as the analysis are available on GitHub\footnote{https://anonymous.4open.science/r/analysis-20F6/DATE/}.
The agent starts the simulation by processing a batch of items and observes for each item a set of metrics, which represent the variables from Table~\ref{tab:model-variables}. In each round, the agent computes the factors that determine its behavior (i.e., $pv$, $ra$, and $ig$) as described, chooses the highest common factor ($cf$), and instructs the engine to operate with the new $batchsize$. This concludes one iteration in the action-perception cycle.



\subsection{Comparative Analysis \& Results}

We evaluated two main aspects of the implementation: (1) which \textit{batch size} it chooses at the end of each cycle, and (2) how well the generative model can reflect the partially observable relation between \textit{utilization} and \textit{part delay}.
%

\begin{figure}[t]
\centering
\subfloat[Batch size / cycle]{\includegraphics[width=0.29\textwidth]{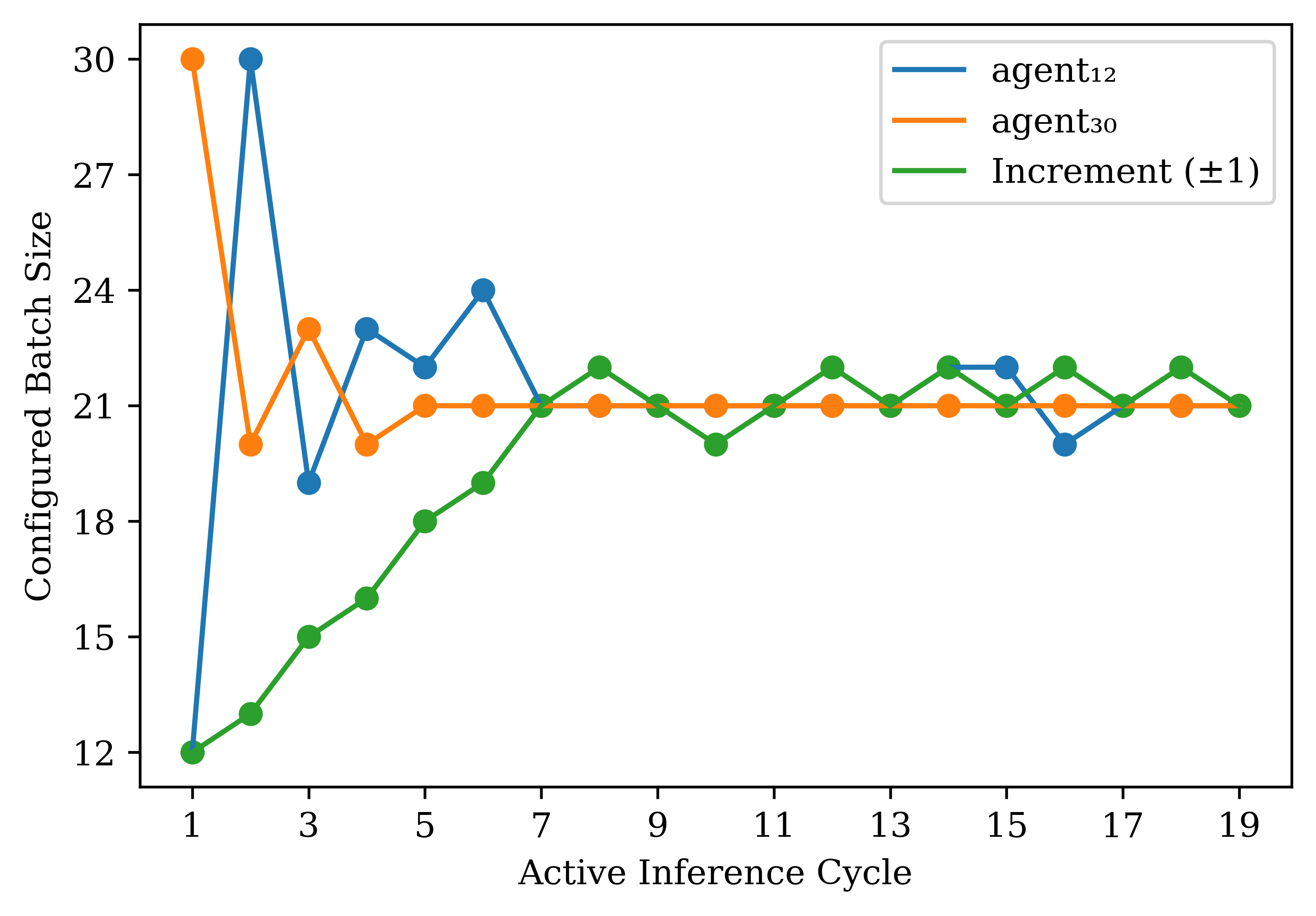}\label{fig:batch-size-history}}
\subfloat[Risk / batch size]{\includegraphics[width=0.21\textwidth]{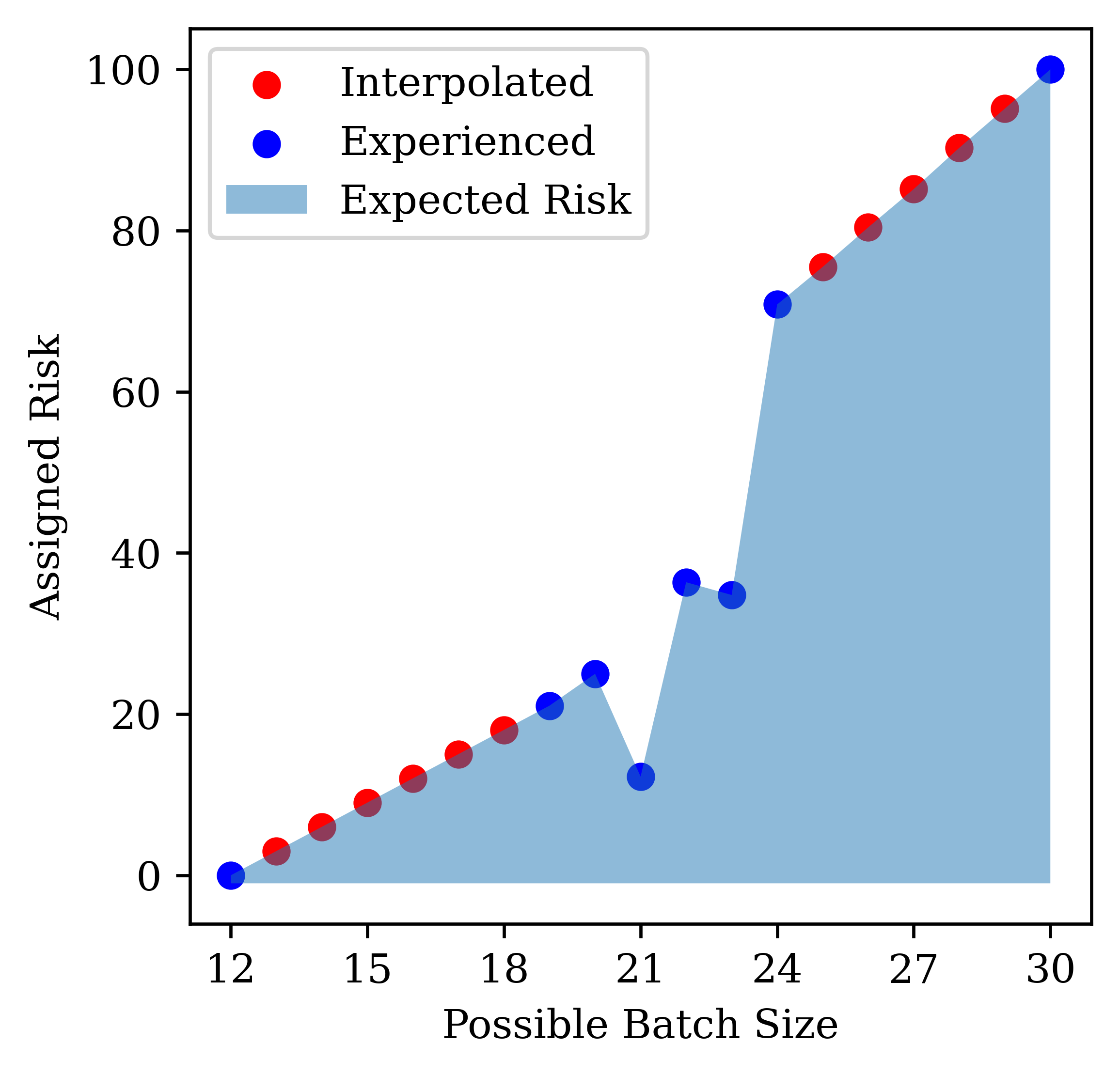}\label{fig:risk-assigned}}
\caption{History of best scoring \textit{batch size} and associated risk}
\label{fig:batch-size-and-risk}
\end{figure}

Fig.~\ref{fig:batch-size-history} tracks the chosen \textit{batch size} depending on the $cf$ score: The blue line depicts the agent's behavior when starting with $batchsize=12$, and the red line when starting with 30, i.e., the safest or most ambitious priors. $Agent_{30}$ reaches a \textit{batch size} of 21 after 5 iterations; whether this is a good (or optimal) solution is determined by multiple opposite factors: As \textit{batch size} increases, both $pa$ and $ra$ rise.
To provide more detail, Fig.~\ref{fig:risk-assigned} contains the $ra$ that $agent_{30}$ assigned to each $batchsize$ after 100 iterations.
Operating with $batchsize=21$, $agent_{30}$ reported SLO violations for 12\% of all observations. If this cannot be tolerated, the $ra$ must be adjusted accordingly; otherwise, 21 presents a very high (if not optimal) $pv$ because any larger \textit{batch size} would be more than three times more likely to violate the SLOs, according to their $ra$. Complementarily, the green line shows the agent's behavior if it simply increases or decreases the \textit{batch size} depending on whether SLOs were fulfilled for the current batch.


\begin{figure}[t]
\centering
\subfloat[Polynomial regression function]{\includegraphics[width=0.321\textwidth]{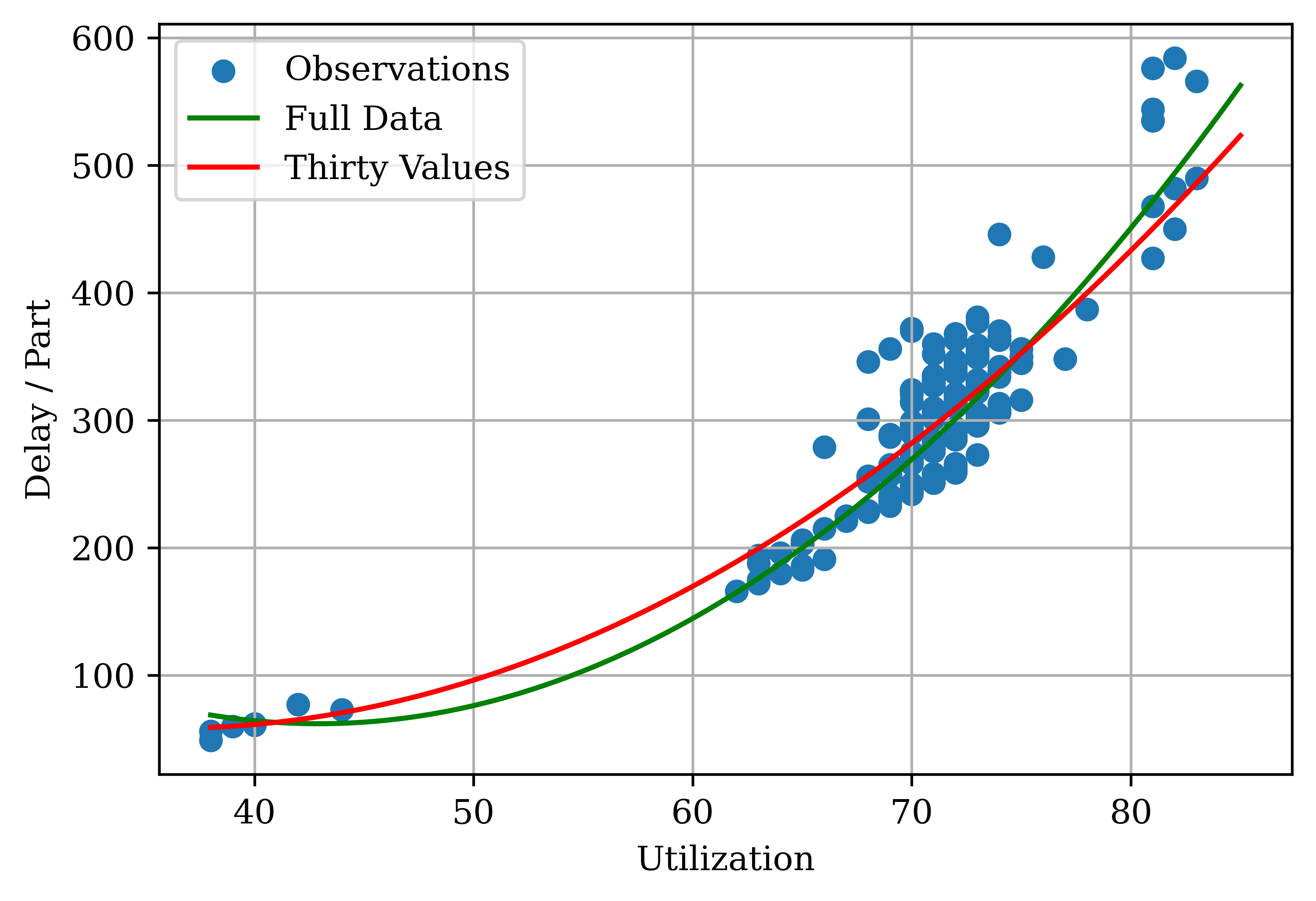}\label{fig:regression}}
\subfloat[Prediction errors]{\includegraphics[width=0.179\textwidth]{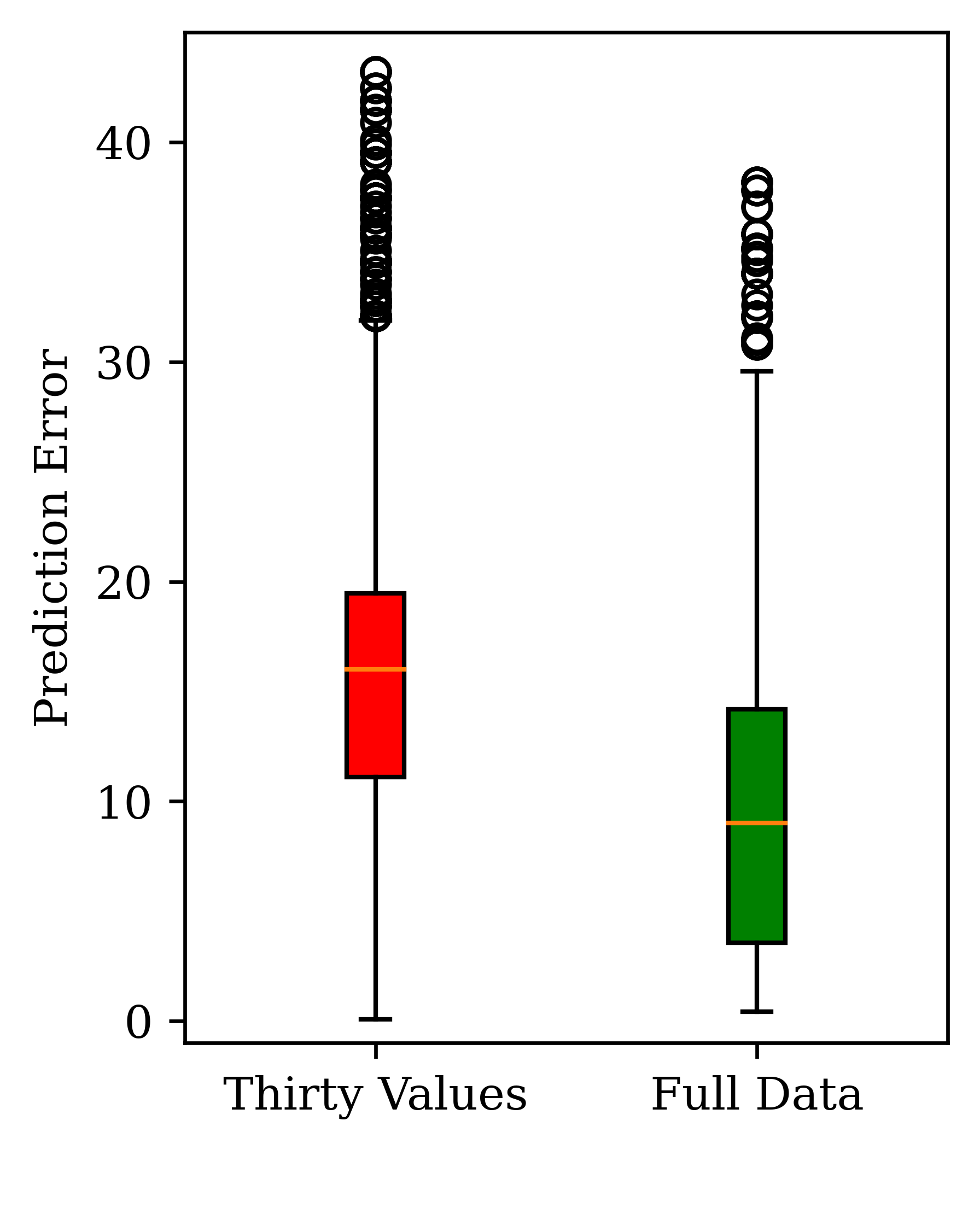}\label{fig:errors}}
\caption{Estimated relation between \textit{utilization} and \textit{part delay}}
\label{fig:estimated-relation}
\end{figure}

While one goal was to reach a high $pv$, the agent's intrinsic motivation is to decrease the FE by developing an accurate generative model. This includes estimating the magnitude of causal relations such as \textit{utilization} $\rightarrow$ \textit{part delay}. Therefore, after receiving a number of samples, the agent can use (polynomial) regression to infer \textit{part delay} for unknown \textit{utilization}. Fig.~\ref{fig:regression} shows a 2D representation of this relation for 2500 processed batch items, supervised by $agent_{12}$; the red line represents the agent's internal model after training on all observations, and the red line after training on only 30 values. After the first round, $agent_{12}$ decided to explore only $batchsize \geq 19$, thus, Fig.~\ref{fig:regression} contains no observations for \textit{utilization} $[45,60]$. The distribution of prediction errors between the regression functions and all items is shown in Fig.~\ref{fig:errors}. We observe that a larger sample size improved the accuracy, but also that a relatively small number of samples (i.e., 30) provided initially acceptable results.

\section{Conclusion}
\label{sec:conclusion}



An ML model should not only reflect a generative process at one moment; ideally, it should persist over time without losing its precision. However, as long as the model is not retrained, a lack of continuous feedback inevitably leads to poor accuracy. Thus, any system that aims to dynamically adapt its service must supervise its inference mechanisms to ensure QoS is fulfilled. With the intention to solve this, we presented a distributed agent that is based on Active Inference -- a concept from neuroscience that describes how the brain constantly predicts and evaluates sensory information to decrease long-term surprise. 
Operating in cycles, the agent maximizes the model evidence by exploring the space of values that fulfill the QoS. Thus, the agent improves any decision-making based on the ML model because ambiguities are repeatedly resolved.

To solve a smart manufacturing use case, we presented a design study that defines the agent's behavior when creating a generative model. Which action the agent takes and how it adapts its beliefs is determined by three main factors: pragmatic value, assigned risk (of violating SLOs), and information gain. We implemented the ACI agent in Python and tracked each cycle's preferred action -- including the factors that led to it -- and the agent's causal understanding between two variables.
After 5 cycles, the agent converged to a solution that presented an optimal tradeoff between high pragmatic value and negligible SLO violations. Further, the agent needed only 30 observations (i.e., 2 cycles) to estimate a previously unknown variable relation. Exploring causalities between variables and constructing the agent's behavior from empirical factors makes the produced solutions traceable. 
Based on these results, we see a strong potential for ACI to support elastic computing systems by continuously ensuring the precision of ML models.

\bibliographystyle{IEEEtran}
\bibliography{references,references_short}

\end{document}